\documentclass{aa}  
\usepackage{graphicx}
\usepackage{subfigure}
\usepackage{txfonts}
\usepackage{verbatim}
\usepackage{float}
\begin{document} 

\title{Asteroseismology of ZZ Ceti stars with fully evolutionary white
  dwarf models} \subtitle{I. The impact of the uncertainties from
  prior evolution on the period spectrum}

\author{F. C. De Ger\'onimo\inst{1,2}, L. G. Althaus\inst{1,2},
  A. H. C\'orsico\inst{1,2}, A. D. Romero\inst{3} and S. O. Kepler\inst{3} }

\institute{$^{1}$ Grupo de Evoluci\'on Estelar y Pulsaciones. Facultad
  de Ciencias Astron\'omicas y Geof\'{\i}sicas, Universidad Nacional
  de La Plata, Paseo del Bosque s/n, 1900 La Plata, Argentina\\ 
  $^{2}$ Instituto de Astrof\'isica La Plata (IALP - CONICET)\\ 
  $^{3}$ Departamento de Astronomia, Universidade
  Federal do Rio Grande do Sul, Av. Bento Goncalves 9500, Porto Alegre
  91501-970, RS, Brazil\\ 
 \email{fdegeronimo@fcaglp.unlp.edu.ar} }
\date{Received ; accepted }

\abstract{ZZ Ceti stars are pulsating white dwarfs with a
  carbon-oxygen  core build up during the core helium burning and
  thermally pulsing Asymptotic Giant Branch  phases. Through the
  interpretation of their pulsation periods by means of
  asteroseismology,  details about their origin and evolution can be
  inferred. The whole pulsation spectrum exhibited by ZZ Ceti stars
  strongly  depends on the inner   chemical structure. At present,
  there are  several processes affecting the chemical 
  profiles that are still  not accurately determined.}{We present a study of the
  impact of the current uncertainties of the white dwarf formation and 
  evolution on
  the expected pulsation properties of ZZ Ceti stars.}  {Our analysis
  is based on a set of carbon-oxygen core white dwarf models with
  masses $0.548$ and $0.837 M_{\sun}$ derived from full evolutionary
  computations from the ZAMS to the ZZ Ceti domain.  We have considered
  models in which we varied the number of thermal pulses,  the amount
  of {\it overshooting}, and the $^{12}$C$(\alpha,\gamma)^{16}$O
  reaction rate within their uncertainties.}{We explore the impact 
  of these major uncertainties
  in prior evolution on  the chemical structure and the expected
  pulsation spectrum.  We find that these uncertainties yield
  significant changes in the  $g$-mode  pulsation periods.}
  {We conclude that the uncertainties in the white
  dwarf  progenitor evolution should be be taken into account in
  detailed asterseismological  analysis of these pulsating stars.} 

\titlerunning{Pulsating white dwarfs}
\maketitle

\section{Introduction}
\label{introduction}

White dwarf (WD) stars are the final result of the evolution of low-
and intermediate-mass \citep[up to $\sim 10.6
  M_{\sun}$,][]{2015ApJ...810...34W} stars. A carbon-oxygen core
composition is expected in most of WDs, which is the result of He
burning in the core of progenitor stars.  The exact proportions of the
resulting carbon-oxygen composition is a key factor that determines
the cooling properties of the emerging WDs. In addition, the precise
shape of the chemical abundance distribution is critical for the
pulsational properties of these stars.

ZZ Ceti (or DAV) variable stars are pulsating WDs  with H-rich
atmospheres. Located in a narrow instability strip with effective
temperatures between 10500 K and 12500 K
\citep{2008PASP..120.1043F,2008ARA&A..46..157W,2010A&ARv..18..471A},
they constitute the most numerous class of compact pulsators. These
stars are characterized by multimode photometric variations caused by
non-radial {\it g}-mode pulsations of low degree ($\ell \leq 2$) with
periods between 70 and 1500 s and amplitudes up to 0.30 mag. These
pulsations are thought to be excited by both the $\kappa-\gamma$
mechanism acting on the base of the H partial ionization zone
\citep{1981A&A...102..375D,1982ApJ...252L..65W} and the ``convective
driving'' mechanism
\citep{1991MNRAS.251..673B,1999ApJ...511..904G,2013EPJWC..4305005S}.
Since the discovery of the first ZZ Ceti star, HL Tau 76 by
\citet{1968ApJ...153..151L}, a continuous effort has been made to
model the interior of these variable stars.

The comparison of the observed pulsation periods in WD with
the periods computed from appropriate theoretical representative
models allows us to infer details about the origin and evolution of
the progenitor star
\citep{2008PASP..120.1043F,2008ARA&A..46..157W,2010A&ARv..18..471A}.
In particular, asteroseismological analysis of ZZ Ceti stars provides
strong constraints on the stellar mass, the thickness of the outer
envelopes, the core chemical composition, and the stellar rotation
rates \citep[e.g.,][]{2012MNRAS.420.1462R}.  In addition, ZZ Ceti
asteroseismology is a valuable tool for studying axions
\citep{Isern92,Isern10,2008ApJ...675.1505B,
  2012MNRAS.424.2792C,2016JCAP...07..036C} and crystallization
\citep{1999ApJ...525..482M,2004A&A...427..923C,2005A&A...429..277C,
  2004ApJ...605L.133M,2005A&A...432..219K,2013ApJ...779...58R}.

There are two main approaches adopted for WD asteroseismology. The
first one employs static stellar models with parametrized chemical
profiles. This approach has the advantage that it allows a full
exploration of the parameter space to find   an optimal seismic model
\citep{2011ApJ...742L..16B,2014ApJ...794...39B,2014IAUS..301..285G,2016ApJS..223...10G}. The
other approach uses fully evolutionary models resulting from the
complete evolution of the progenitor star, from the Zero Age Main Sequence
(ZAMS) to the WD stage
\citep{2012MNRAS.420.1462R,2013ApJ...779...58R}. This approach
involves the most detailed and updated input physics, in particular
regarding the internal chemical structure, a crucial aspect for
correctly disentangle the information encoded in the pulsation
patterns of variable WDs.  This method has been successfully applied
in different classes of pulsating WDs \citep[see][in the case of ZZ Ceti 
stars]{2012MNRAS.420.1462R,2013ApJ...779...58R}.
However, asteroseismological inferences based on this approach do not
take  into account the current uncertainties neither
in the modeling nor in the
input physics of WD progenitors. An assessment of the impact
of these uncertainties on the pulsation periods of ZZ Ceti stars,
constitutes the core feature of the present paper.
 
In this regard, there exist several important uncertainties linked to
WD prior evolution that should be explored. As well known,
the exact proportions of the carbon-oxygen composition of the WD core
constitute a key factor that strongly impact the cooling properties of
the emerging WDs. In addition, the precise shape of the chemical
abundance distribution is critical for the pulsational properties of
WDs. The main uncertainty affecting the carbon-oxygen chemical profiles and the
cooling times  in WDs is related with the
$^{12}$C$(\alpha,\gamma)^{16}$O reaction rate.
\citet{2010ApJ...716.1241S} studied the differences found in cooling
times that arise from uncertainties in several processes during the
pre-WD evolution, and found that the changes in cooling times
associated with the $^{12}$C$(\alpha,\gamma)^{16}$O reaction rate was
of about 7\%. More recently, \citet{2016ApJ...823...46F} have carried
out an extensive study of the impact of uncertainties in several
reaction rates during core H and He burning on important stellar
parameters as mass, age, carbon-oxygen central abundances and central
density. They found that uncertainties in the
$^{12}$C$(\alpha,\gamma)^{16}$O and $^{14}$N$(p,\gamma)^{15}$O are
dominant. Another uncertain process that modifies the chemical
structure is overshooting (OV). The precise size of the
convective core and the mixing regions are currently one of the major
uncertainties affecting the computations of stellar structure and
evolution. In particular, the amount of OV considered during core He
burning is not completely known and strongly modifies the internal 
carbon-oxygen
profile, the mass of the convective core and the temperature
stratification.  Current uncertainties  in physical processes
operative during the  thermally pulsing Asymptotic Giant Branch
(TP-AGB) phase also leave their signature in the final chemical
abundance distribution expected in  WDs. In particular, the
outermost-layer chemical stratification of the WD core is built up during
the TP-AGB phase and depends strongly on the  occurrence of OV and
mass loss during this stage. In particular, the mass of the intershell
rich in He and C that is left by the  short-lived He flash convective
zone depends on both the amount of OV and the number of thermal pulses
(TP) which is determined by the poorly  constrained efficiency of mass loss
\citep{2014PASA...31...30K}. 

In this paper, we will assess the impact of the main uncertain  processes in the
evolutionary history of the progenitors  on the chemicla profiles
and the pulsation period spectrum of ZZ Ceti stars.  In particular, 
we  focus on the effects
of OV during the core He burning stage, the occurrence and
number of TP in the AGB phase, and the uncertainties in
the $^{12}$C$+\alpha$ reaction rate.  To this end, we computed
evolutionary sequences from the ZAMS, through the TP-AGB stage
following the evolution of the remnant star to the WD state until the
domain of the ZZ Ceti instability strip. The sequences are
characterized by progenitor masses of 1.5 and $4 M_{\sun}$ with
initial metallicity of $Z= 0.01$.

This paper is organized as follow: in Sect. \ref{cap:tools} we
introduce the numerical tools employed and the input physics assumed
in the evolutionary calculations, and the pulsation code employed.
In Sect.\ref{cap:chemical-profiles-pulsation} we present our results.
Finally, in Sect. \ref{cap:conclusions} we conclude the paper by
summarizing our findings.


\section{Computational tools}
\label{cap:tools}

\subsection{Evolutionary code and input physics}

The DA WD evolutionary models calculated in this work were
generated with the {\tt LPCODE} evolutionary code, which produces
detailed WD models based on updated physical description
\citep{2005A&A...435..631A,2010ApJ...717..897A,
2010ApJ...717..183R,2012MNRAS.420.1462R,2016A&A...588A..25M}.  {\tt
  LPCODE} has been employed to study different aspects of the
evolution of low-mass stars \citep{2011A&A...533A.139W,
  2013A&A...557A..19A,2015A&A...576A...9A}, formation of  Horizontal
Branch stars \citet{2008A&A...491..253M},  Extremely Low Mass WDs
\citep{2013A&A...557A..19A}, AGB and post-AGB evolution
\citep{2016A&A...588A..25M}. We detail below the main physical
ingredients of {\tt   LPCODE} relevant to this work:

\begin{itemize}

\item standard mixing-length theory with the free parameter $\alpha =
  1.61$  was adopted for pre-WD stages; 

\item diffusive OV during the evolutionary stages prior to
  the TP-AGB phase, were allowed to occur following the description of
  \citet{1997A&A...324L..81H}. We adopted $f= 0.016$ for all
  sequences, except when indicated. OV  is relevant for the
  final chemical stratification of the WD
  \citep{2002ApJ...581..585P,2003ApJ...583..878S};

\item breathing pulses at the end of core He burning effect,  which
  occur at the end of core He burning, were suppressed  (see
  \citet{2003ApJ...583..878S} for a discussion on this topic);

\item {\tt LPCODE} considers a simultaneous treatment of
  non-instantaneous mixing and burning of elements.  Our nuclear
  network accounts explicitly for the following 16 elements: $^1$H,
  $^2$H, $^3$He, $^4$He, $^7$Li, $^7$Be, $^{12}$C, $^{13}$C, $^{14}$N,
  $^{15}$N, $^{16}$O, $^{17}$O, $^{18}$O, $^{19}$F, $^{20}$Ne,
  $^{22}$Ne and 34 thermonuclear reaction rates
  \citep{2005A&A...435..631A};

\item gravitational settling, and thermal and chemical diffusion was
  taken into account for $^1$H, $^3$He, $^4$He,$^{12}$C,$^{13}$C,
  $^{14}$N,$^{16}$O \citep{2003A&A...404..593A};

\item during WD phase, chemical rehomogenization of the inner carbon-oxygen
  profile induced by Rayleigh-Taylor (RT) instabilities was
  implemented following \citet{1997ApJ...486..413S};

\item for the high-density regime characteristic of WDs, we
  used the equation of state of \citet{1994ApJ...434..641S}, which
  accounts for all the important contributions for both the liquid and
  solid phases.
  
\end{itemize}

We computed the evolution of our sequences throughout all  the
evolutionary stages of the WD progenitor, including the stable  core
He burning and the TP-AGB and post-AGB phases.  We want to mention
that {\tt LPCODE} has been compared against other WD
evolution code showing that the differences in the WDs
cooling times that comes from the different numerical implementations
of the stellar evolution equations were by $2 \%$
\citet{2013A&A...555A..96S}.

\subsection{Pulsation code}
\label{pulsation-codes}

For the pulsation analysis presented in this work 
we employed the {\tt LP-PUL} adiabatic non-radial
pulsation code described in \citet{2006A&A...454..863C}, coupled to
the {\tt LPCODE} evolutionary code. {\tt LP-PUL} solves the full
fourth-order set of real equations and boundary conditions governing
linear, adiabatic and non-radial stellar pulsations.  
The pulsation code provides
the eigenfrequency $\omega_{\ell, k}$ ---where $k$ is the radial order
of the mode--- and the dimensionless eigenfunctions
$y_1, \cdots, y_4$. {\tt LP-PUL} computes
the periods ($\Pi_{\ell, k}$), rotation splitting coefficients 
($C_{\ell, k}$), oscillation kinetic energy ($K_{\ell, k}$) 
and weight functions ($W_{\ell, k}$). The expressions to compute
these quantities can 
be found in the Appendix of \citet{2006A&A...454..863C}.
The Brunt-V\"ais\"al\"a frequency 
($N$) is computed as \citep{1990ApJS...72..335T}:

\begin{equation}
N^2 =\frac{g^2 \rho}{P}\frac{\chi_{\rm T}}{\chi_{\rm P}}(\nabla_{\rm ad}-\nabla + B),
\label{eq:brunt}
\end{equation}

\noindent where

\begin{equation}
B= -\frac{1}{\chi_{\rm T}}\sum_{i=1}^{n-1} \chi_{ X_{\rm i}} \frac{d \ln X_{\rm i}}{d \ln \rho},
\label{eq:ledoux}
\end{equation}

\noindent is the Ledoux term and contains the contribution 
coming from the chemical composition changes, and

\begin{equation}
\chi_{\rm T} = \left [\frac{\partial \ln P}{\partial T} \right ]_{\rho},~ 
\chi_{\rho} = \left [\frac{\partial \ln P}{\partial \rho} \right ]_{\rm T},~ 
\chi_{X_i} = \left [\frac{\partial \ln P}{\partial X_i} 
\right]_{\rho, {\rm T}, X_{j \neq i}}.
\end{equation}

\begin{figure*}[ht]
  \centering
  \mbox{
    \subfigure[\label{pt-perfil-0548}]{\includegraphics[width=6cm, angle=270]{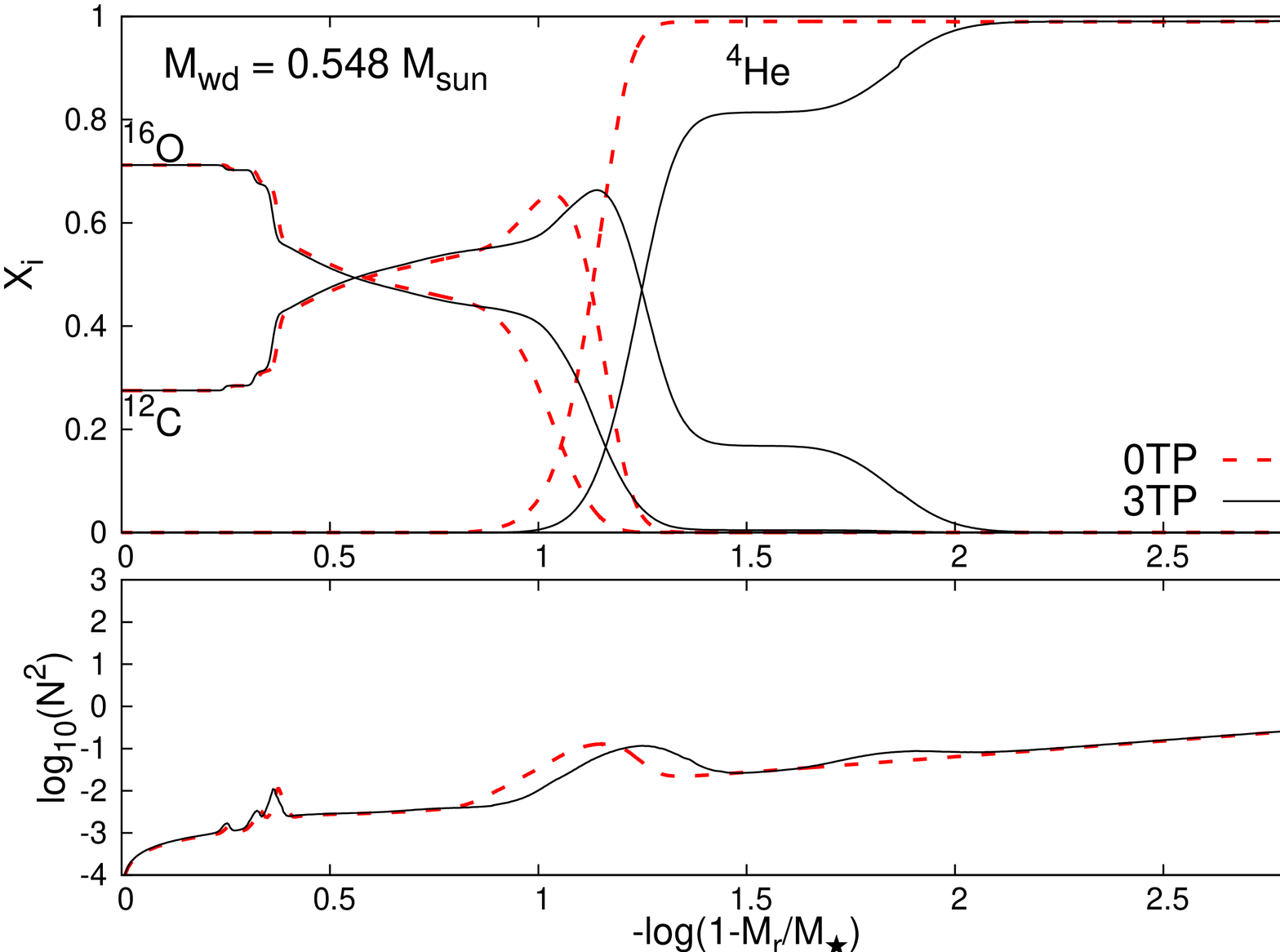}}\quad
    \subfigure[\label{pt-perfil-0837}]{\includegraphics[width=6cm, angle=270]{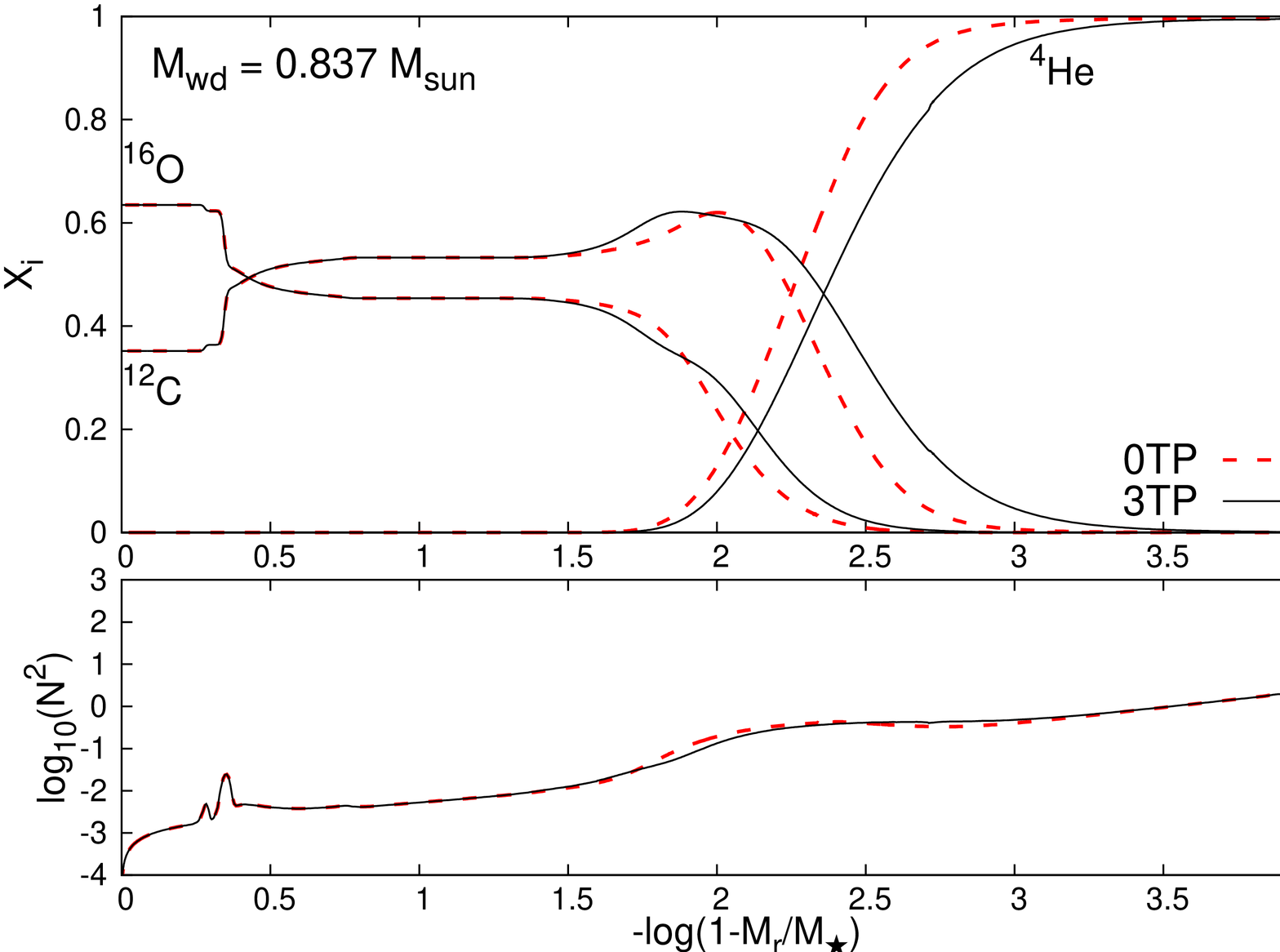}}\quad
    
  }
  \caption{Upper panels: inner O, C and He abundance distribution in
    terms of the outer mass fraction for the two stellar models 
    considered in our analysis at $T_{\rm eff}\sim 12\,000$ K. 
    0TP referes to the expected chemical profiles resulting from a 
    progenitor before the occurrence of the first TP (red dashed line). 
   3TP corresponds to the case of the chemical profiles resulting from a 
   progenitor at end of the third TP (black solid line).} 
  \label{pt-perfil}
\end{figure*}

\begin{figure*}[ht]
  \centering
  \mbox{
    \subfigure[\label{pt-perfil-0548}]{\includegraphics[width=5.8cm, angle=270]{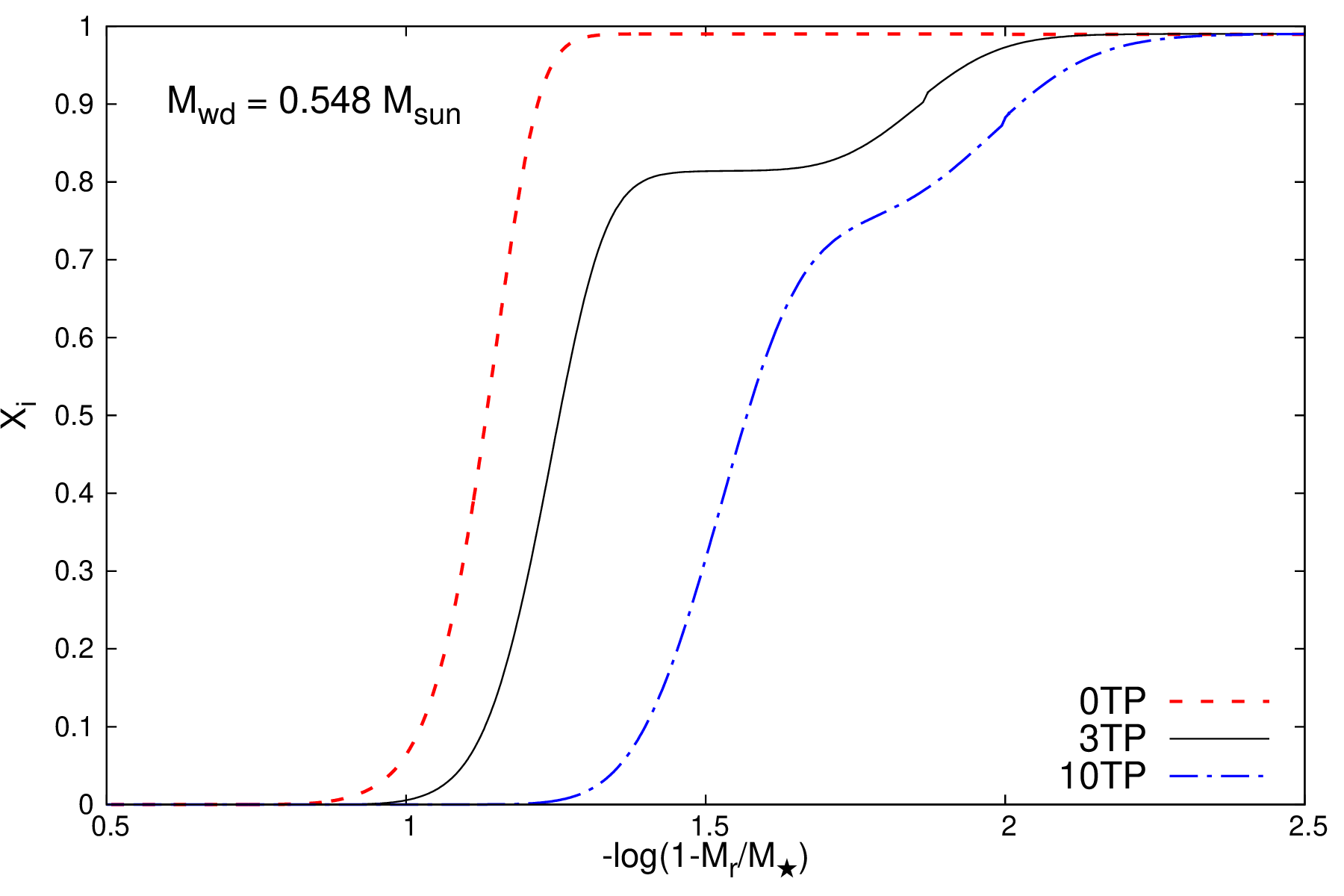}}\quad
    \subfigure[\label{pt-perfil-0837}]{\includegraphics[width=5.8cm, angle=270]{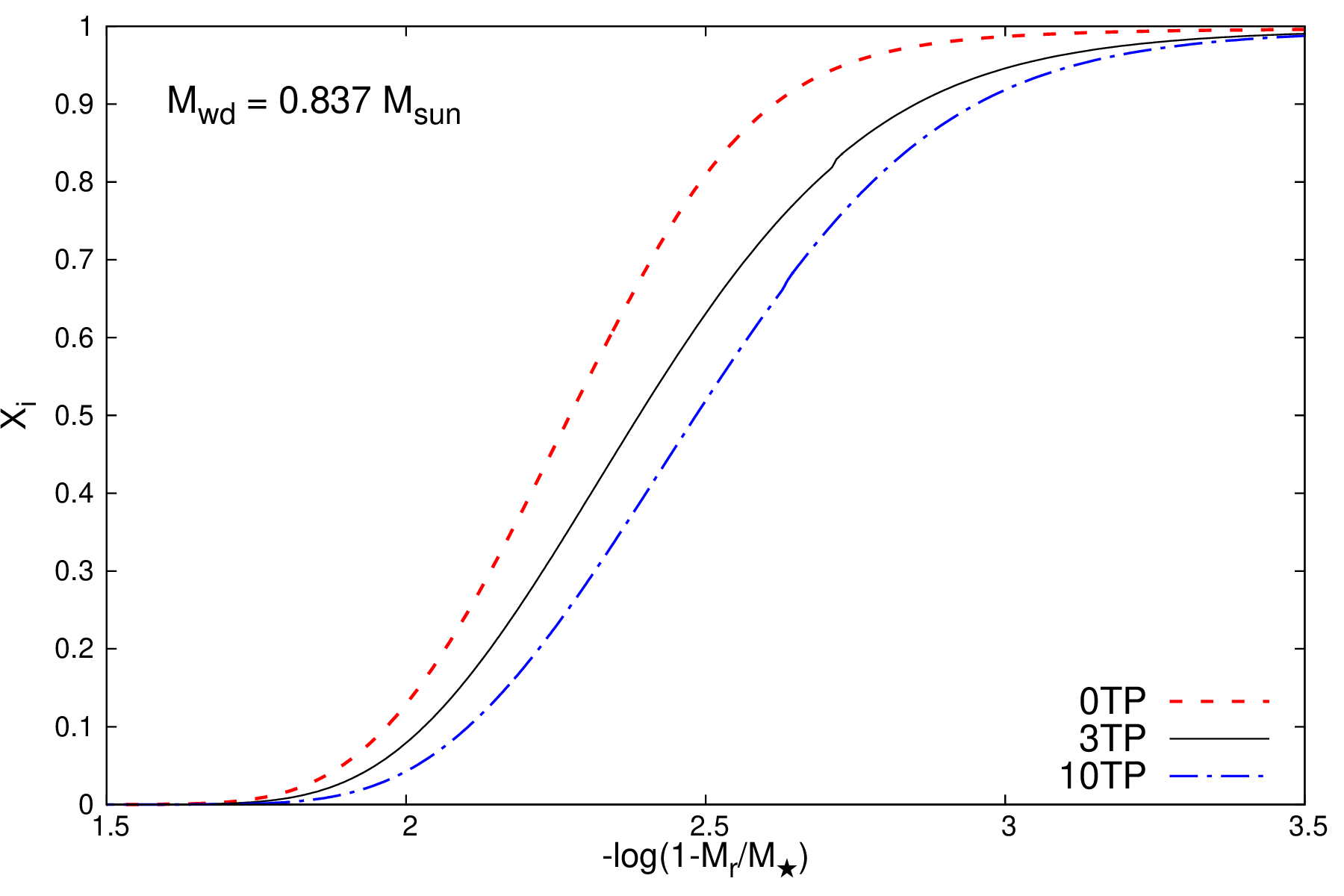}}\quad
    
  }
  \caption{Inner He abundance distribution in terms of the outer
    mass for models at the  ZZ Ceti stage for which the progenitor 
 experienced no TP 
(0TP), three TP (3TP), and ten TP (10TP) models. The intershell region 
is eroded completely for the massive model. }
  \label{fig:pt-helio}
\end{figure*}

\begin{figure*}[ht]
  \centering
  \mbox{
    \subfigure[\label{pt-Pdif-0548}]{\includegraphics[width=6cm, angle=270]{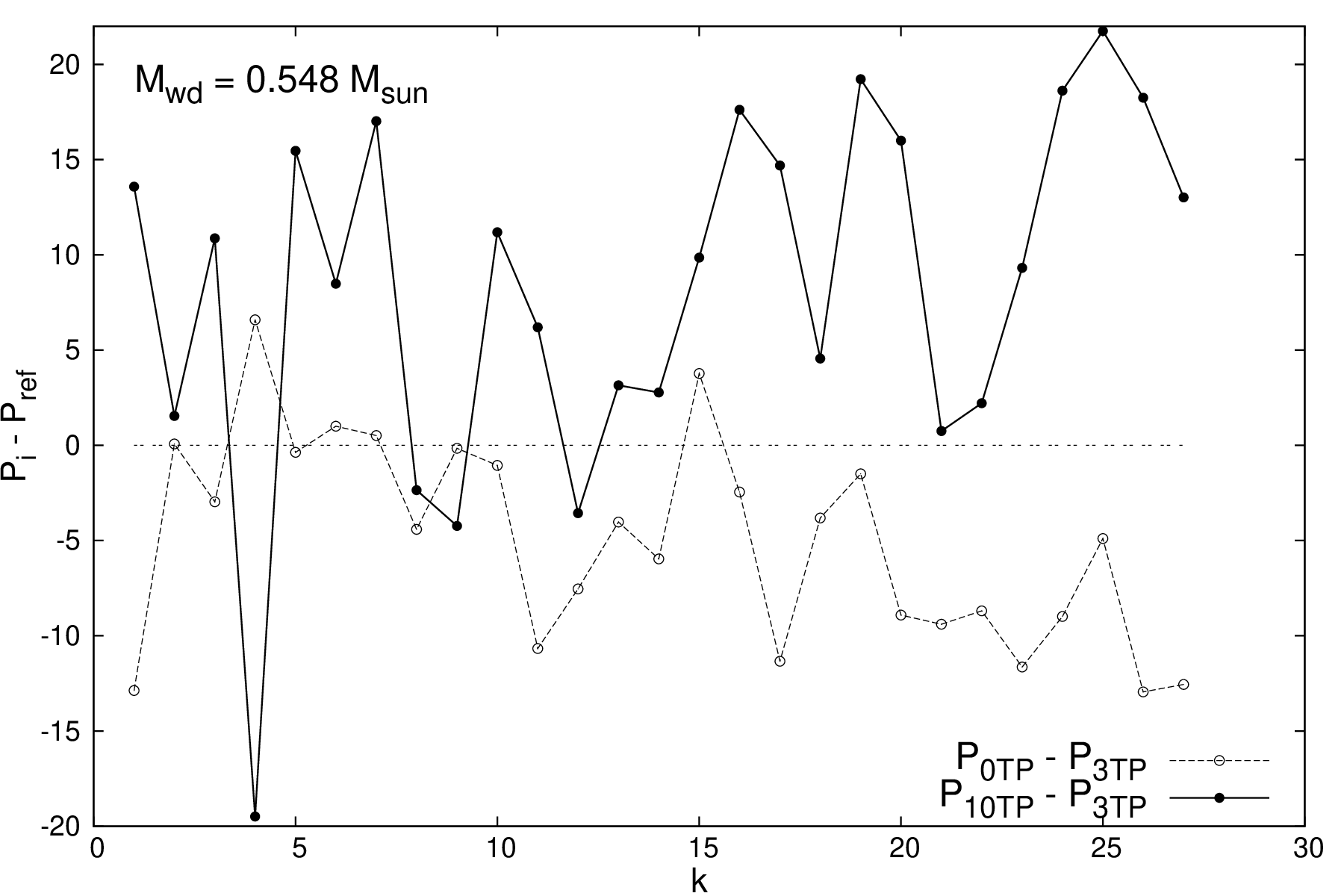}}\quad
    \subfigure[\label{pt-Pdif-0837}]{\includegraphics[width=6cm, angle=270]{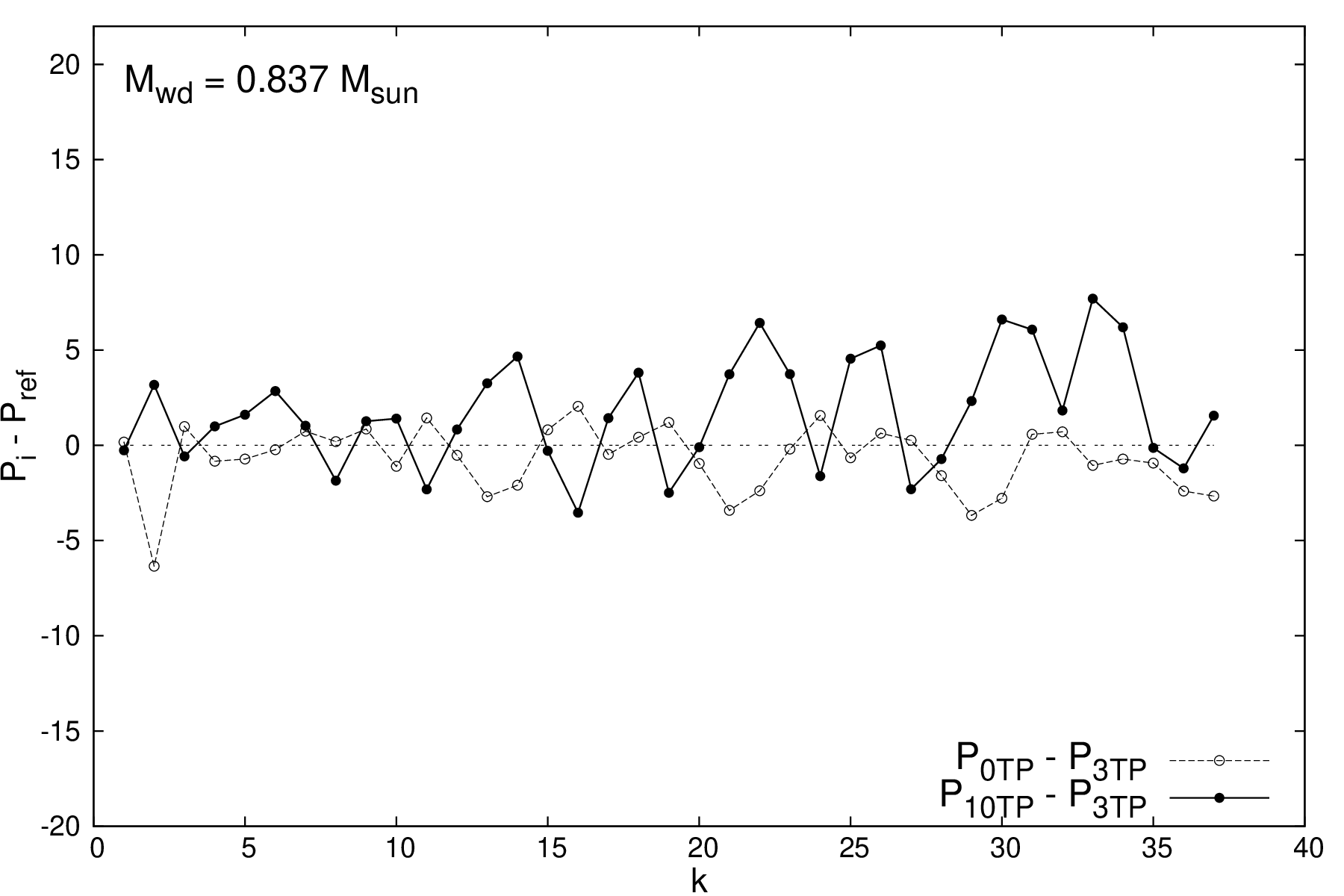}}\quad
    
  }
  \caption{Period differences (with fixed radial order $k$) in terms of $k$ between
    the 0TP and 3TP models (black dashed line) and 
    between the 10TP and 3TP models (black solid line) for the 
    $M_{\star}= 0.548 M_{\sun}$ and  $M_{\star}= 0.837 M_{\sun}$ WD models
   (left and right panels, respectively).}
  \label{pt-Pdif}
\end{figure*}

\begin{figure*}[ht]
\centering
  \mbox{
    \subfigure[\label{ov-perfil-0548}]{\includegraphics[width=6cm, angle=270]{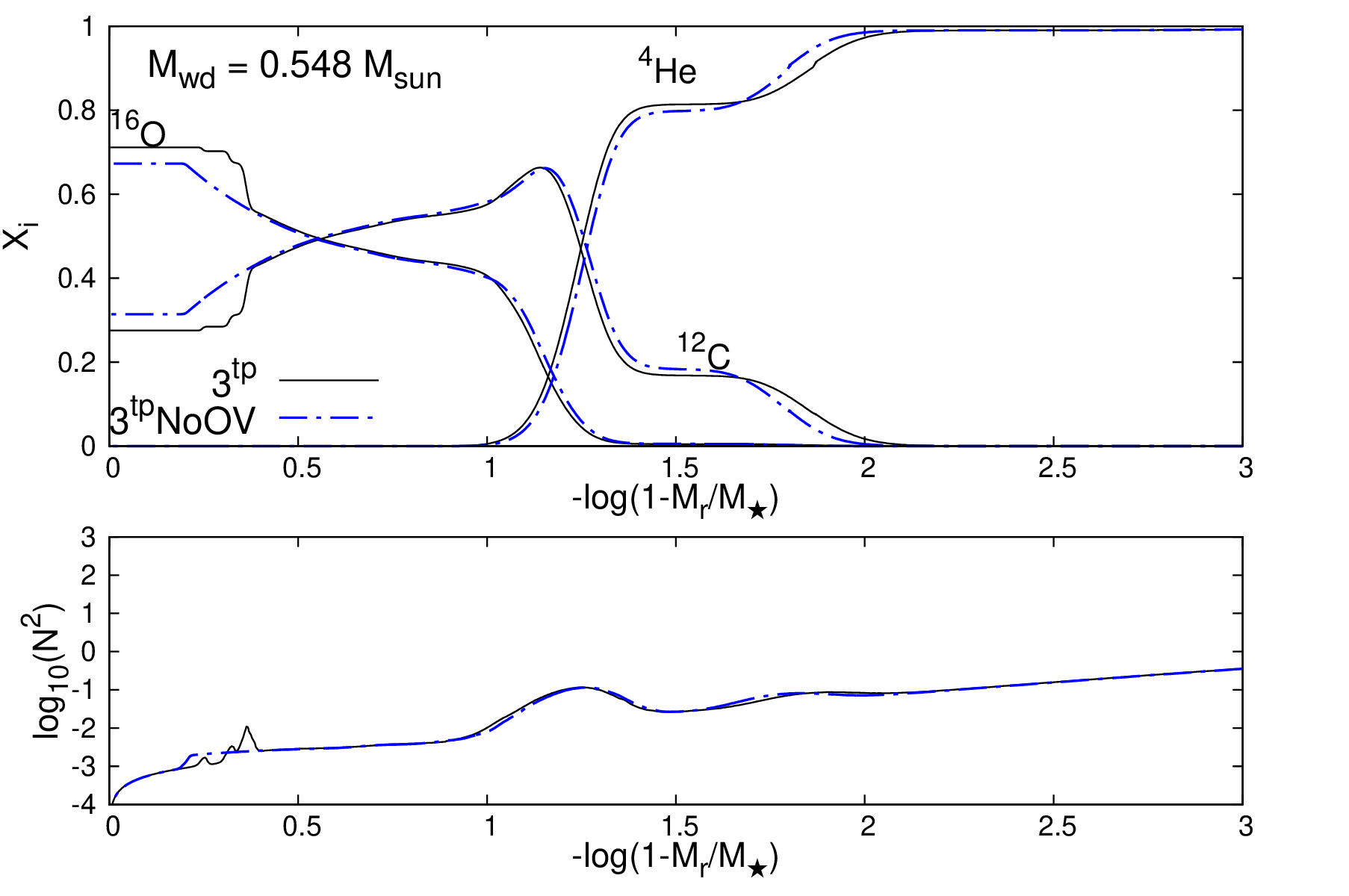}}\quad
    \subfigure[\label{ov-perfil-0837}]{\includegraphics[width=6cm, angle=270]{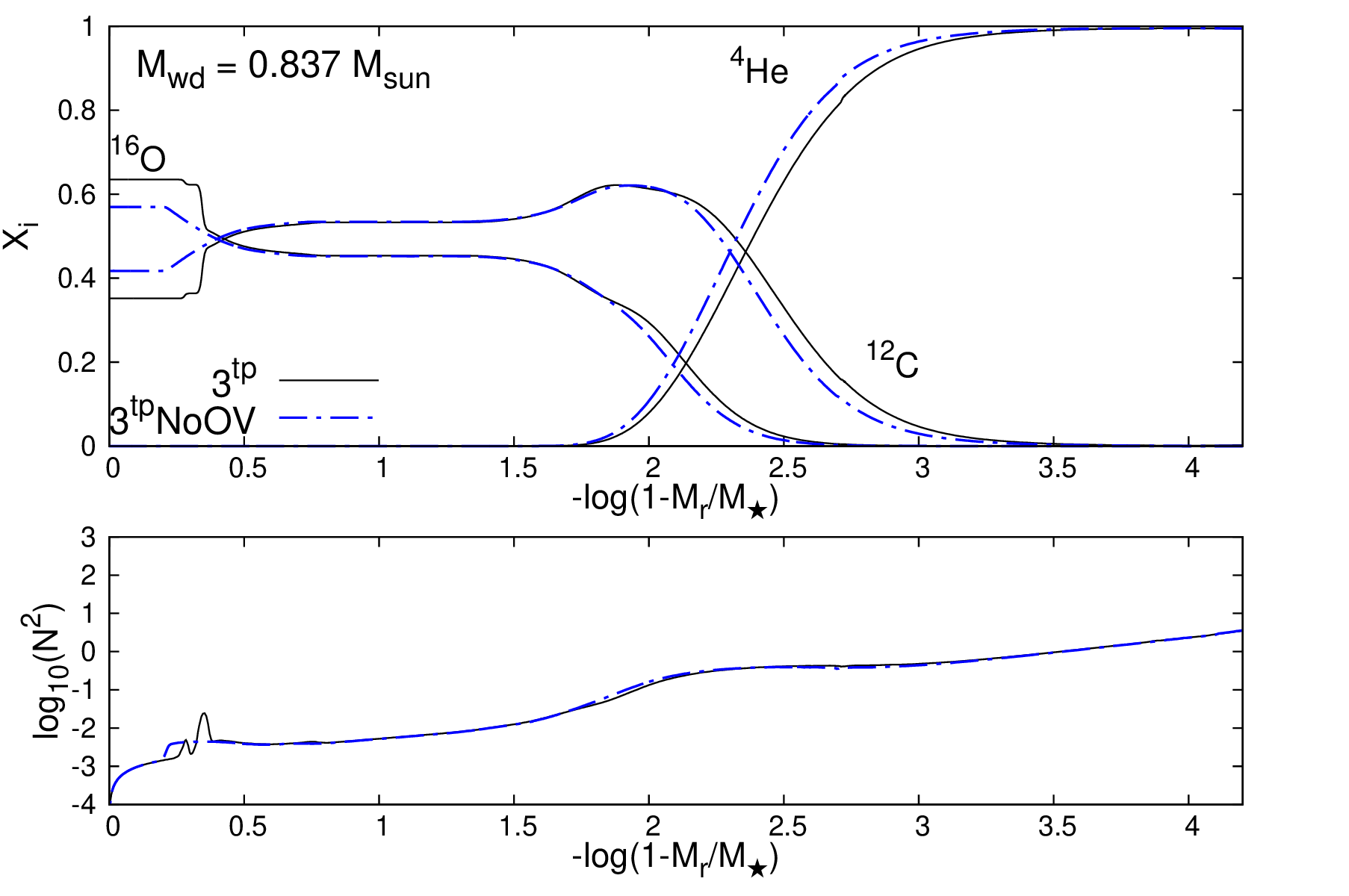}}\quad
    
  }
  \caption{Same as Fig. \ref{pt-perfil}, but for models at the third TP
  calculated with an OV parameter $f= 0.016$ (3TP) and with $f= 0$
    (3TP-NOV) during core He burning.}
  \label{ov-perfil}
\end{figure*}

\begin{figure*}[ht]
\centering
  \mbox{
    \subfigure[\label{ov-Pdif-0548}]{\includegraphics[width=6cm, angle=270]{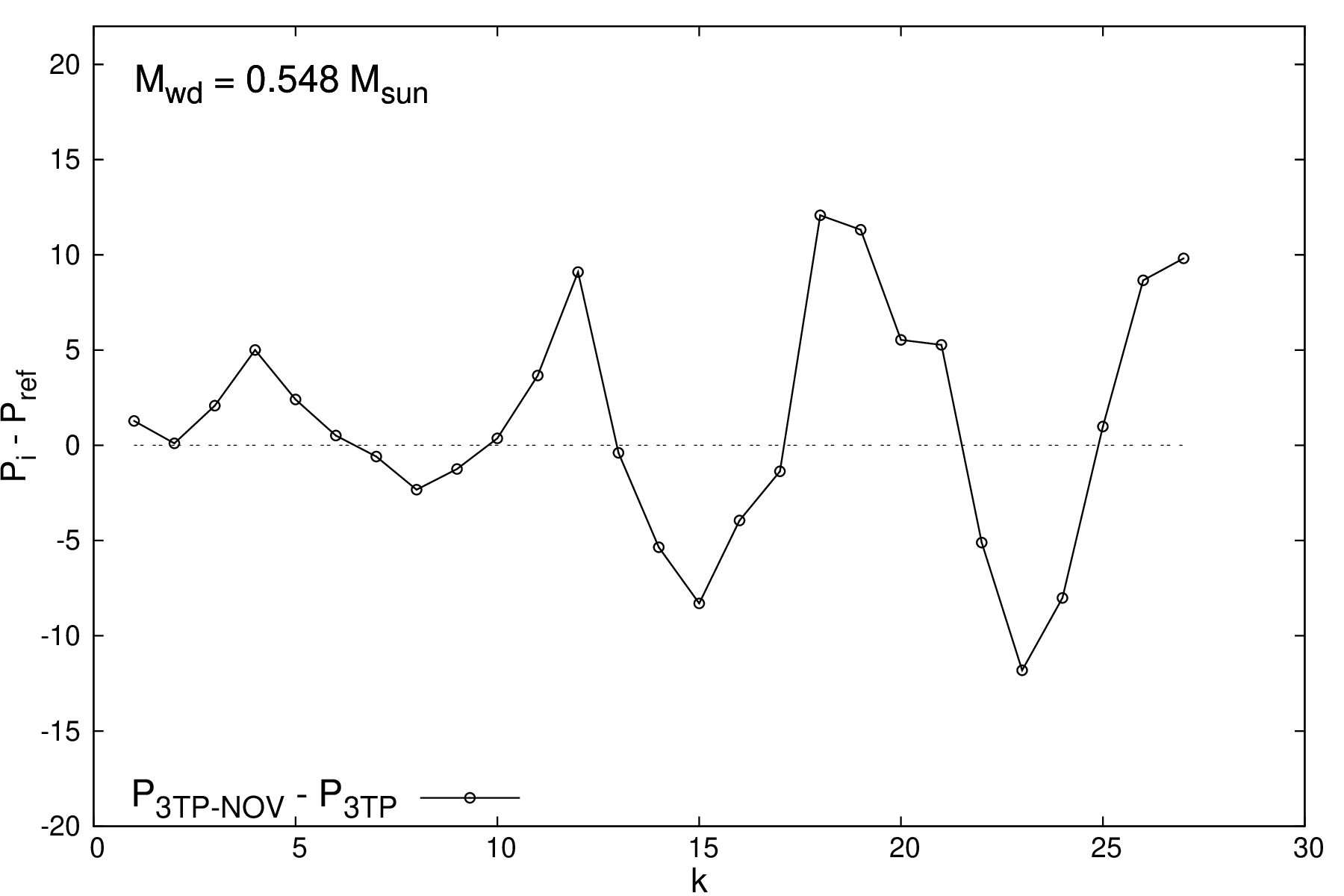}}\quad
    \subfigure[\label{ov-Pdif-0837}]{\includegraphics[width=6cm, angle=270]{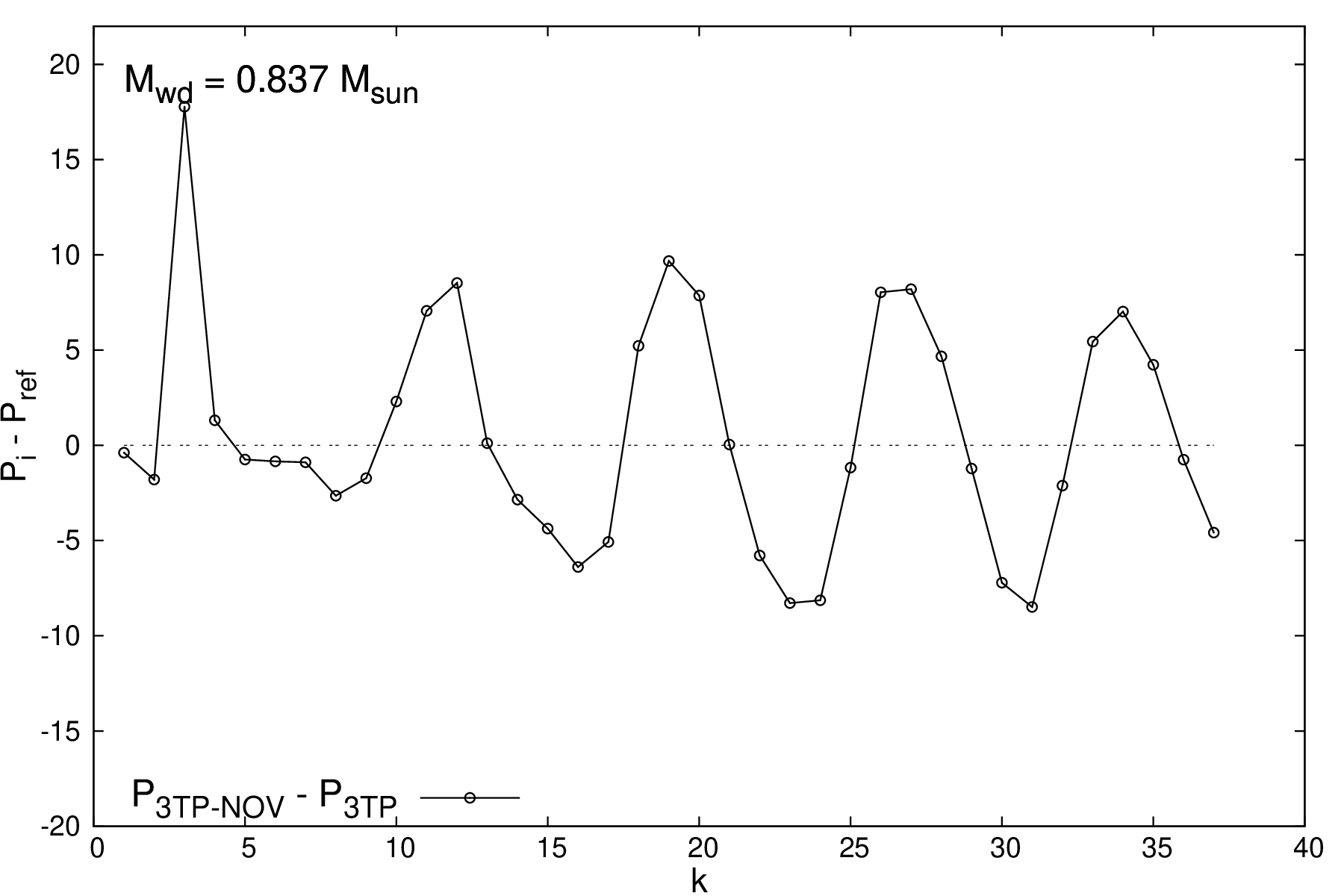}}\quad
    
  }
  \caption{Same as Fig. \ref{pt-Pdif} but for the 3TP and
    3TP-NOV models shown in Fig.\ref{ov-perfil} .}
  \label{ov-Pdif}
\end{figure*}


\section{Pulsational analysis}
\label{cap:chemical-profiles-pulsation}

In this section we study the uncertainties during the prior evolution
of the progenitor star that affects the chemical structure of the 
emerging WDs and their pulsation period spectra. In particular, we explore:

\begin{enumerate}
\item The occurrence of TP during the AGB phase of the
progenitor star. \label{it:TP}
\item The occurrence of OV during core He burning. \label{it:OV}
\item The $^{12}$C$(\alpha,\gamma)^{16}$O reaction rate. \label{it:CO}
\end{enumerate}

We concentrate on template models 
characterized by $M_{\star}= 0.548, 0.837 M_{\sun}$,
with H envelope mass of $M_{\rm H} \sim 4\times 10^{-6} M_{\star}$
at $T_{\rm eff} \sim 12\,000$ K, and considered the pulsation period spectrum for 
modes with $\ell= 1$.

In Fig. \ref{pt-perfil} we show the internal chemical profiles and the
logarithm squared Brunt-V\"ais\"al\"a frequency in terms of the outer 
mass fraction for both
models at the ZZ Ceti domain.  Our models are
characterized by three chemical transition regions from center to the
surface: a chemical interface of C and O, a double-layered 
chemical structure involving He, C and O, and an interface of 
He/H. The existence of these three transition zones 
induces ``bumps'' in the shape of the Brunt-V\"ais\"al\"a frequency
which strongly affects the whole structure of the pulsation spectrum
of the star, in particular the mode-trapping properties
\citep{1992ApJS...80..369B,1996ApJ...468..350B,2002A&A...387..531C}. 
These features play a crucial role acting
as ``filters'' by trapping the mode oscillation energy close to the
surface, or confining it in the core.  

Next, we compare the pulsation properties of the models
considered for each case. Since the models compared have the same
stellar mass \footnote{We rescaled the stellar mass when neccesary.}, 
effective temperature and mass of the H envelope, the
differences found in the pulsation properties are strictly due to the
differences in the chemical structure resulting from different kinds of 
the progenitor evolution.

\subsection{Occurrence of TP in AGB}

In this section we explore the impact on the chemical profiles when
the progenitor evolves through the TP-AGB phase. Because during this
phase a pulse-driven convection zone develops, the building of an
intershell region and a double-layered chemical structure is expected
at the bottom of the He buffer. The size of this region decreases 
with the number of TP, i.e., as the star evolves through the TP-AGB, as
consequence of the He-burning shell. The number of TP
experienced by the progenitor star is uncertain and depends on the rate at 
which mass is lost during the TP-AGB phase, on the initial metallicity
of progenitor star, as well as, to a less extent,  on the occurrence
of extra-mixing in the pulse-driven convection zone 
\citep{2000A&A...360..952H}. 
To quantify the impact of all these uncertainties on the pulsational 
properties of ZZ Ceti stars, we have explored the situation in which the
progenitor experiences 3 and 10 TP (hereinafter referenced as 
3TP and 10 TP models) by forcing the departure from the TP-AGB 
phase at specific stages. In addition, we have considered the extreme 
situation in which the star is forced to abandon the AGB phase before the
occurrence of the first TP (hereinafter referred to as 0TP model).

In Fig. \ref{pt-perfil} we display the chemical profiles (upper panel)
for the most abundant elements, and the logarithm of the squared 
Brunt-V\"ais\"al\"a
(B-V) frequency (lower panel) in terms of the outer mass. As mentioned, the
evolution through the TP-AGB leaves its signature on the chemical
profile of our models with the formation of an intershell region rich
in He and C at the bottom of the He buffer. The presence
of C in the intershell region stems from the short lived
convective mixing that has driven the C-rich zone upward during
the peak of the He pulse on the AGB
\citep{2010ApJ...717..897A}. As we mentioned, the size of this 
intershell region
depends on the number of TP experienced by the star, decreasing
as the number of TP increases (see later in this Section).  For the
low-mass models, we find that this intershell region formed at the
TP-AGB is expected to survive when the stars reach the ZZ Ceti
phase. On the other hand, for more massive stars we do not expect this
region to be present at the ZZ Ceti stage.  This is due that the
diffusion processes acting during the WD cooling phase are more
efficient for massive stars, thus completely eroding it
\citep{2010ApJ...717..897A}.  We note that the impact on the chemical
structure, and in consequence, on the B-V frequency, is more
pronounced for the less massive model (see lower panels of
Fig.\ref{pt-perfil}).  The results for the 0TP model are shown with 
dashed red lines in the upper panels of Fig. \ref{pt-perfil}. 
Note that in this case, no intershell region is expected. In the case of 
the 10 TP model sequence, the intershell region becomes less massive, 
and it is removed by diffusion
by the time the ZZ Ceti stage is reached more easily than in the case
of the 3TP model. This is borne out by Fig. \ref{fig:pt-helio} which
depicts the He abundance distribution below the envelope of the
emerging WD.

The impact of considering different number of TP on the period
spectrum of the models, can be seen in Fig. \ref{pt-Pdif}. In this
plot, we show the resulting period differences ($\Delta P_{k} \equiv
P_{k, {\rm 0TP}} -P_{k, {\rm 3TP}}$) between the 0TP and 3TP models
(solid line, filled circles) and the period differences ($\Delta P_k
\equiv P_{k, {\rm 10TP}} -P_{k, {\rm 3TP}} $) between the 10TP and 3TP
models (dashed line, empty circles) as a function of the radial order
$k$. We found that the resulting changes in the periods (at fixed $k$)
are more pronounced in the case of the less massive model. These
  substancial variations ---high as 22 s--- are not solely due to the
  intershell region (or the overall shape of the profile), but also due
  to the shift outward of the core/He main chemical transition region.

Not surprisingly, the expectation for the massive model are markedly
different. Indeed, in this case  the period differences are on average 
between 2 and 3 s, with some modes reaching period differences 
as high as $\sim 8$ s.  These smaller changes resulting in the case of the 
massive model are due to the slight  differences found in chemical profiles. 
In fact, no intershell is expected to occur at the  ZZ Ceti domain.

\begin{figure}[ht] 
\begin{center}
  \includegraphics[width=9cm, angle=0]{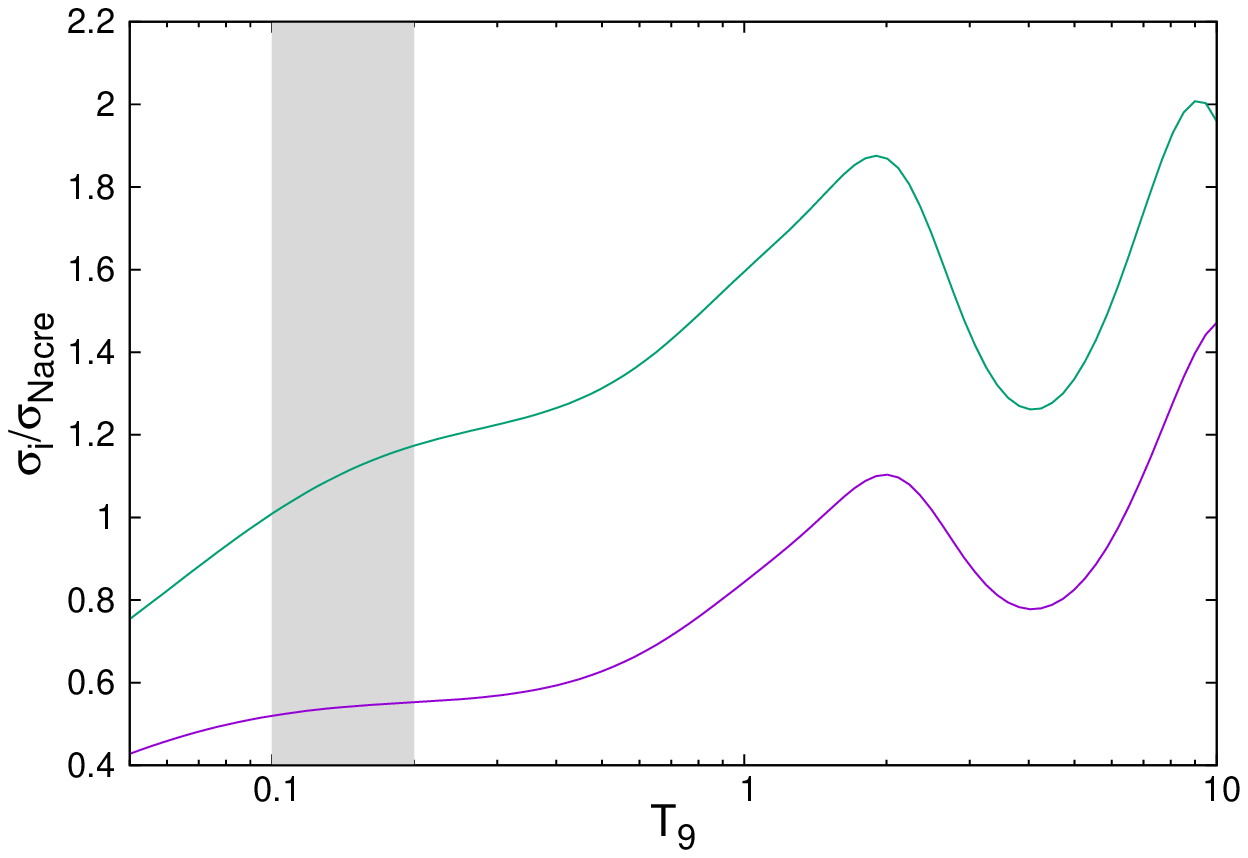}
  \caption{Ratio between the Kh and Nacre, and Kl an Nacre reaction 
rates (upper and 
lower curves, respectively) as a function of temperature. Grey region denotes 
the  typical temperatures expected during the core-He burning.}
  \label{fig:rates}
\end{center}
\end{figure}

\subsection{Occurrence of OV during core He burning}

The occurrence of extra mixing episodes during core He burning
strongly affects the final shape of the carbon-oxygen core.  To quantify the
impact of the occurrence of core OV on the period spectrum
of ZZ Ceti stars, we have considered the extreme situation in which no
OV is allowed during the core He burning (3TP-NOV
model), and compared it to the template 3TP model, which, 
as we mentioned, takes into account OV during core He burning.

We start by examining Fig. \ref{ov-perfil} that shows the inner
chemical profiles at the ZZ Ceti stage. 
Note that models with OV are characterized by lower central C
abundances. This is because the ingestion of fresh He in a C-rich zone
favours the consumption of C via the reaction $^{12}$C$+\alpha$, with
the consequent increase of $^{16}$O.  It is worth noting that this
change in the central value of C and O does not modify the B-V
frequency (lower panel). However, the B-V frequency is notoriously
affected by the steep variation in the carbon-oxygen profile at
$-\log(1-Mr/M_{\star}) \sim 0.4$ left by core OV.  The
resulting differences in the value of the periods are shown in
Fig. \ref{ov-Pdif}.  We find that OV introduces absolute differences 
of 2 to 4 s on average, markedly lower than the
period differences resulting from uncertainties in the TP-AGB phase.
Note also that the average of the period differences resulting from
OV are similar for both stellar masses.

\subsection{Uncertainties in the $^{12}$C$(\alpha,\gamma)^{16}$O reaction
  rate}

\begin{figure*}
  \centering \mbox{
    \subfigure[\label{co-perfil-0548}]{\includegraphics[width=6cm,
        angle=270]{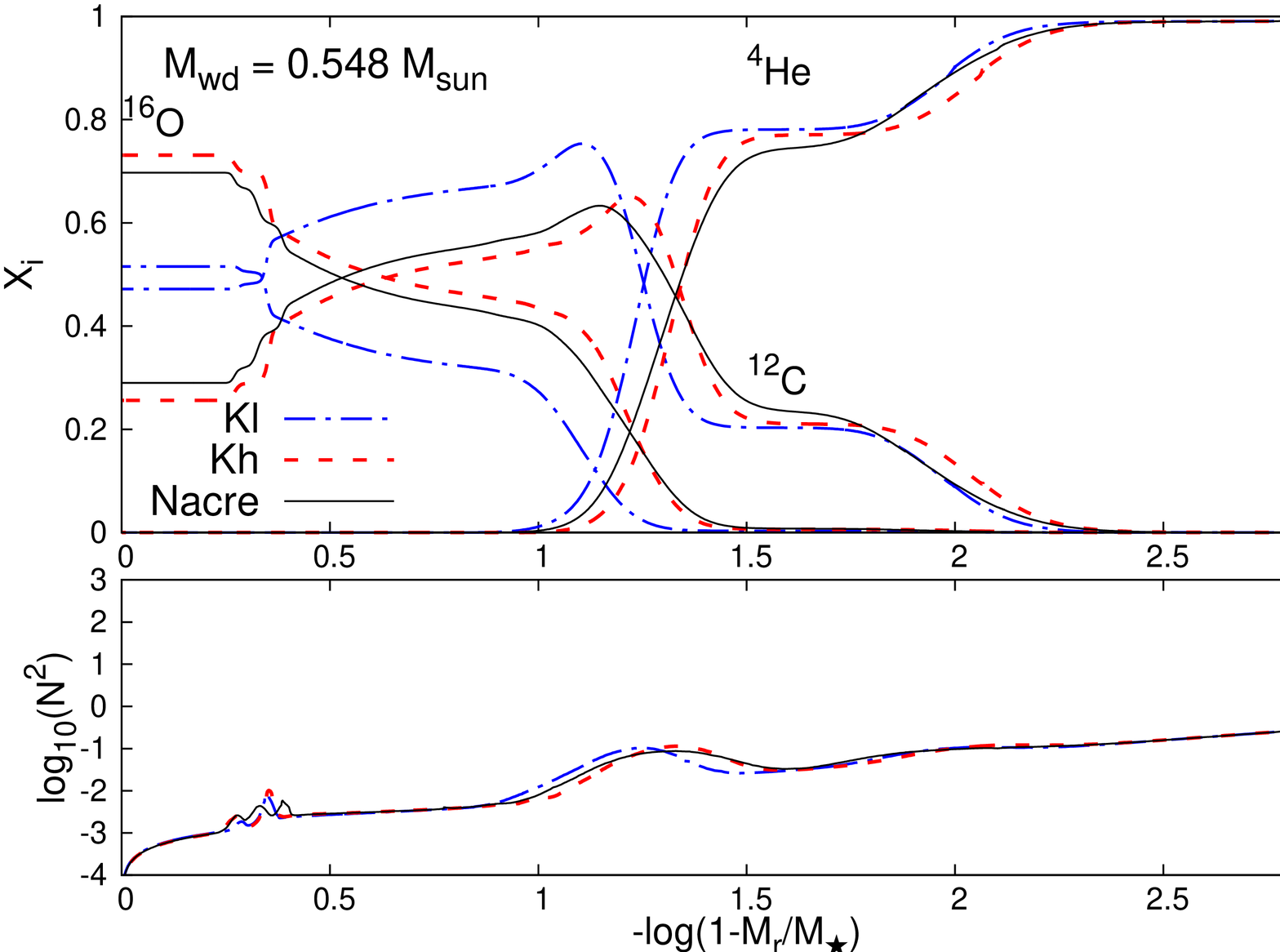}}\quad
    \subfigure[\label{co-perfil-0837}]{\includegraphics[width=6cm,
        angle=270]{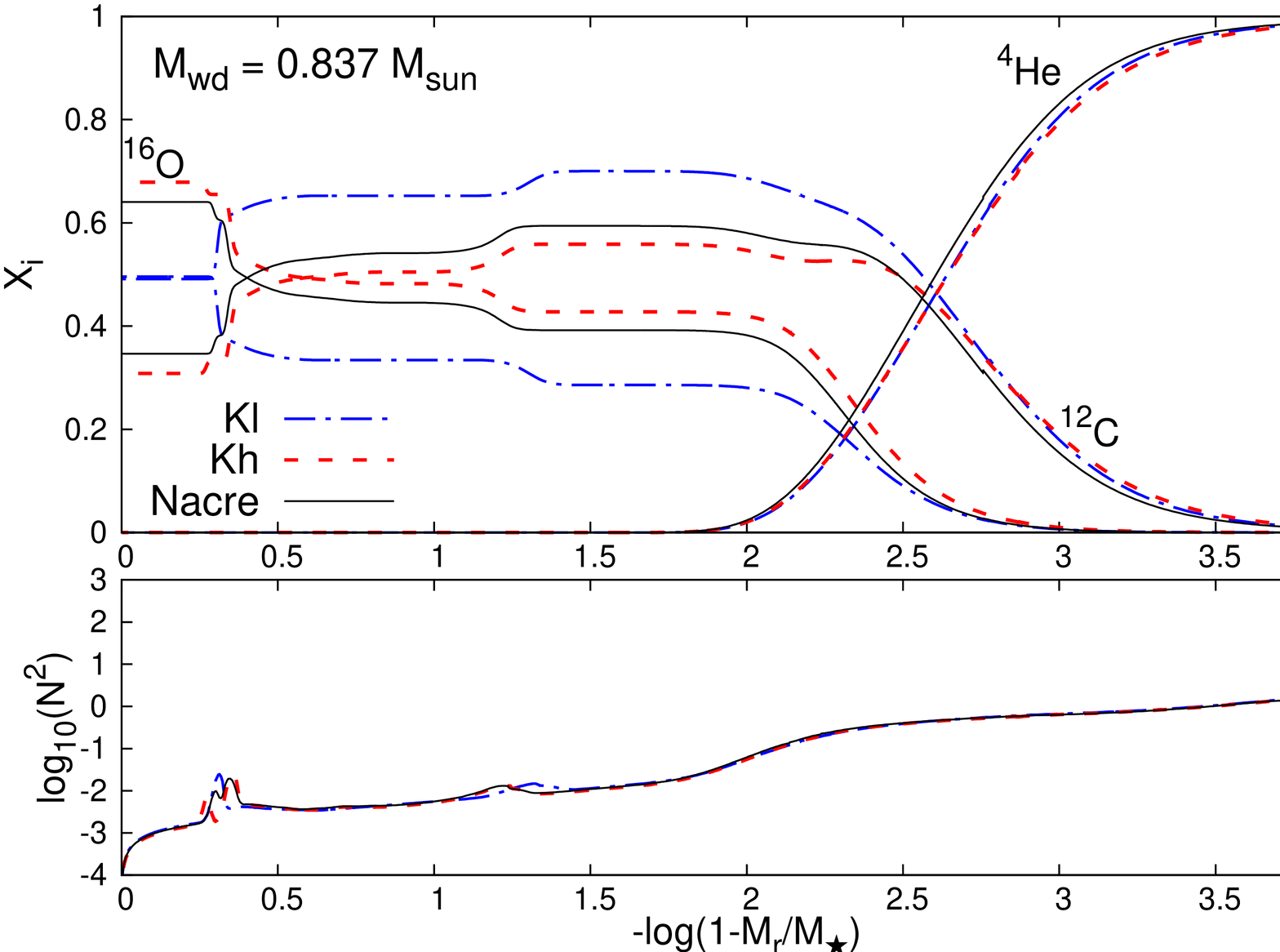}}\quad
  }
  \caption{Same as Fig. \ref{pt-perfil} but for models computed
  with three different $^{12}$C$+\alpha$ reaction rates:
  \citet{1999NuPhA.656....3A} (Nacre, black solid line), the higher
  (KH), and the lower (KL) reaction rates  from \citet{2002ApJ...567..643K} (red
  dashed line and blue dashed line, respectively).
}
  \label{co-perfil}
\end{figure*}

To determine the impact of the uncertainties in the $^{12}$C$+\alpha$
reaction rate, we computed the core He burning phase of the
progenitors by considering three different values for the
$^{12}$C$(\alpha,\gamma)^{16}$O reaction rate, and followed the
evolution until the progenitor experiences several TP in the AGB-TP
phase.  We considered the reaction rates provided by
\citet{1999NuPhA.656....3A} (Nacre, our reference model), and the
extreme high and low values from \citet{2002ApJ...567..643K} (Kh and
Kl models, respectively). These high and low values are derived from
the uncertainty for the recommended rate listed in that work.  The use of
these extreme values for this reaction rate allows us to account for
the uncertainties in the carbon-oxygen chemical profiles of WD. In
Fig.  \ref{fig:rates} we plot the temperature dependence of the 
ratio between the Kh and Nacre, and Kl an Nacre reaction rates. Note that
for core He burning temperatures ($\sim 2\times10^8$ K, grey
zone), $\sigma_{\rm Kl}/ \sigma_{\rm Nacre}\sim 0.55$ and $\sigma_{\rm
  Kh}/ \sigma_{\rm Nacre}\sim 1.1$.  The impact on the chemical
profiles at the ZZ Ceti stage is depicted in Fig. \ref{co-perfil}.  We
found differences in the carbon-oxygen abundances, in the location of the O/C
and O/C/He chemical transitions, and in the abundances of the
intershell region. As expected, the $^{16}$O abundance is larger for
the models computed with the Kh reaction rate.  Models that have been
computed assuming the Nacre and Kh reaction rates exhibit quite
similar chemical structures. Differences in the B-V frequency are 
found at the carbon-oxygen and triple transition
(see lower panels of Fig. \ref{co-perfil}).

\begin{figure*}
  \centering \mbox{
    \subfigure[\label{co-Pdif-0548}]{\includegraphics[width=6cm,
        angle=270]{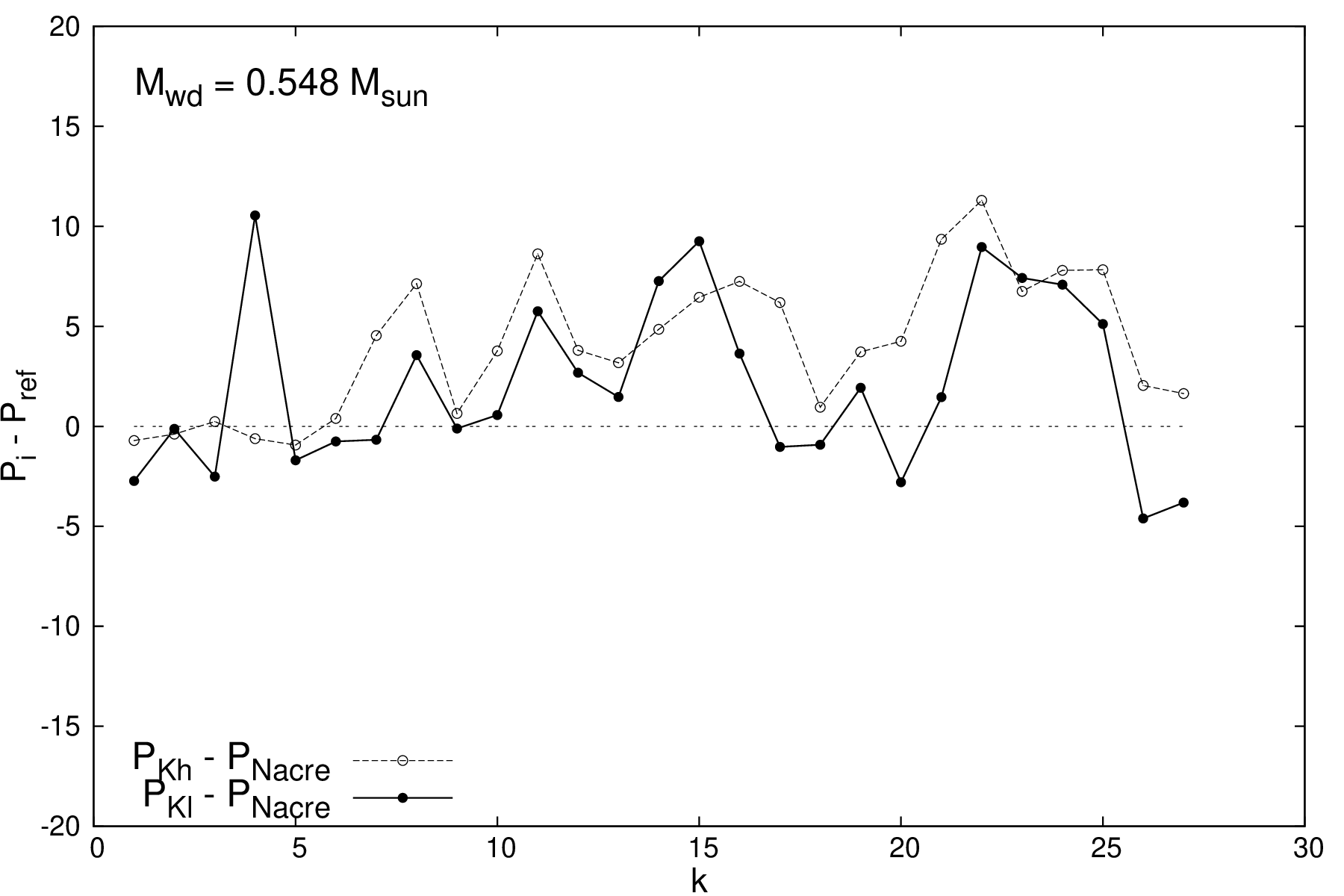}}\quad
    \subfigure[\label{co-Pdif-0837}]{\includegraphics[width=6cm,
        angle=270]{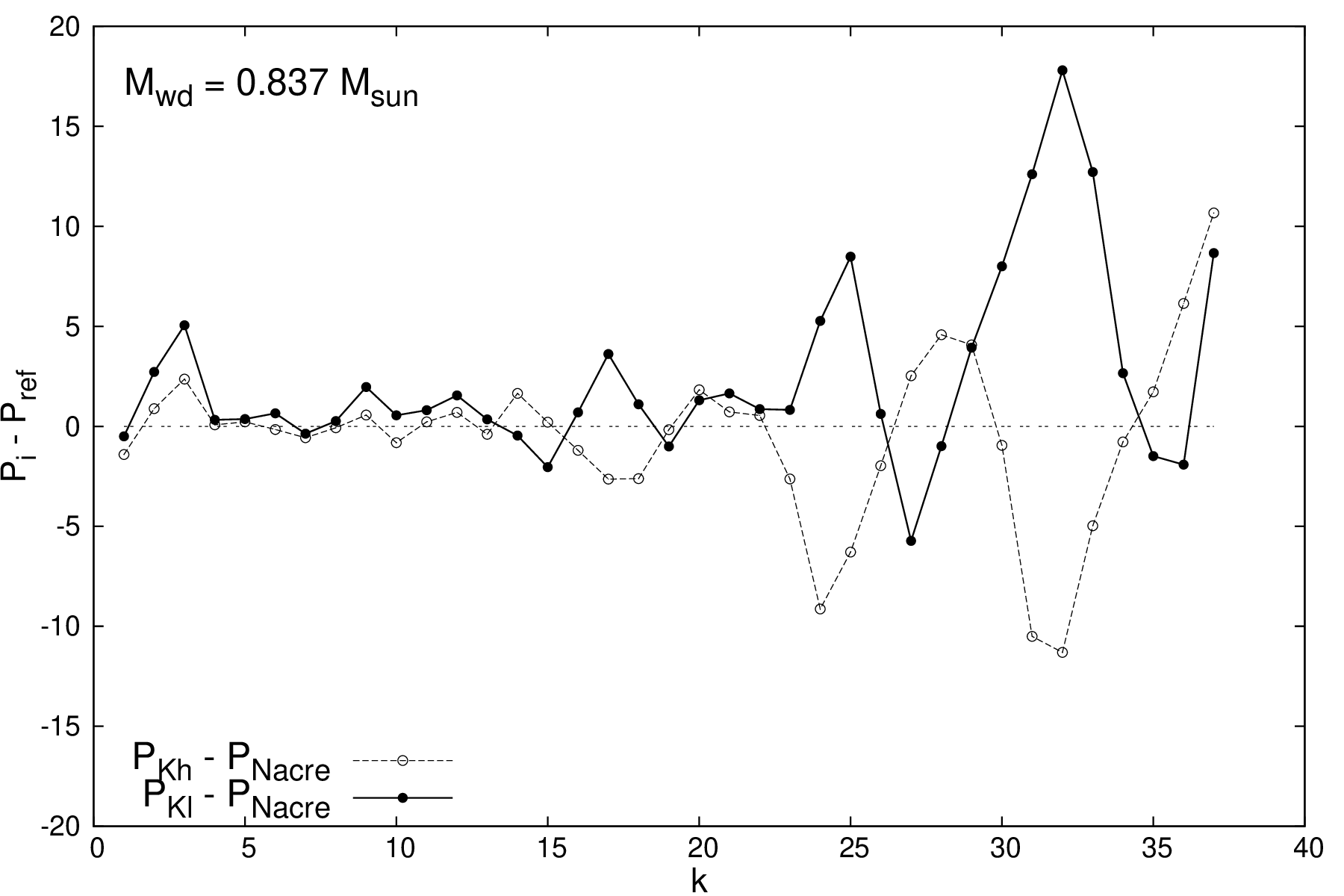}}\quad
    
  }
  \caption{Same as Fig. \ref{pt-Pdif} but for the period differences
between  Kh and Nacre models,
    (black dashed line) and Kl and Nacre models (black solid line).  }
  \label{co-Pdif}
\end{figure*}

The resulting impact on the pulsation periods can be seen in
Fig. \ref{co-Pdif}, where we plot the differences between  
the periods predicted by the Kh and Nacre models
(dashed line, empty circles) and the differences predicted by the 
Kl and Nacre models (solid line, filled
circles). Low-order modes are, in general, more affected when
the Kl reaction rate is assumed. In particular, for the 
$0.548 M_{\sun}$ model, appreciable differences are expected 
for both low and high radial-order modes, with maximum differences of
$|\Delta P|\sim 11$ s.  By constrast, in the case of the massive model, 
small period differences are found for modes with radial orders $k < 23$
independently of the adopted rate, with a maximum of $|\Delta P|\sim
5$ s. These small variations found in low-order modes reflect the
much more smoother behaviour of the chemical profile at the outermost
part of the carbon-oxygen core in the case of the massive model. 
But, for radial orders $k > 23$, appreciable differences are found, reaching
$|\Delta P|\sim 18$ s at most. The appreciable
differences found for modes with radial orders $k > 23$
stems from the fact that some modes of high radial order $k$ are 
sensitive to the chemical structure of the core. 

\section{Summary and conclusions}
\label{cap:conclusions}

In this paper, we studied the impact of uncertainties from progenitor
evolution on the pulsational properties of ZZ Ceti WD stars.
We focused on the occurrence of TP in the AGB stage, the
occurrence of  OV during the core He burning and  also the
uncertainties in the  $^{12}$C$(\alpha,\gamma)^{16}$O reaction
rate. Our models were derived from the full and complete evolution
from the ZAMS, through the TP-AGB. 
Evolution of our model sequences  was followed  to the domain of
ZZ Ceti stars, for which we  computed the adiabatic $g$-mode pulsation
properties.   We computed different sets of models in which we varied
the amount of  OV during core He burning, the
$^{12}$C$(\alpha,\gamma)^{16}$O  reaction rate, and the occurrence and
number of TP during the  TP-AGB phase.

We have found that if progenitor evolves or not through the TP-AGB
phase makes an important point as far as the pulsational properties at
the ZZ Ceti stage are concerned. In particular, we found that the
resulting changes in the periods (at fixed $k$) are more pronounced in
the case of the less massive model, with period differences on average
between 5 and 10 s. However, for massive models, the period
differences are markedly smaller ($\sim 2-3$ s). This is mainly
  due to the shift of the core/He main chemical transition is less pronounced
  and the double layered chemical structure build up during the last
  TP on the AGB, which impacts the period spectrum of pulsating WDs
  \citep{2010ApJ...717..897A}, is completely eroded by element
  diffusion in the massive WDs ($M_{\star} \gtrsim 0.80 M_{\sun}$) by
  the time the domain of the ZZ Ceti stars is reached.

On the other hand, the occurrence of OV during the core He 
burning leaves imprints on the Brunt-V\"ais\"al\"a 
frequency, resulting in absolute differences of 2 to 4 s on average in the 
pulsation periods, as compared with 
the situation in which OV is disregarded. This conclusion is 
the same regardless the stellar mass.

Finally, uncertainties related with the
$^{12}$C$+\alpha$ reaction rate affects both O, C, and He abundances and the
location of the chemical transitions. In the case of less massive models, 
the changes in period induced by
these different chemical structures amount to, at most, 
$|\Delta P|\sim 11$ s for some specific modes. For our massive model,
period differences are relevant only for $k \geq 20$ modes. We mention 
that the period differences resulting from uncertainties of the 
$^{12}$C$+\alpha$ reaction rate are less relevant than the period 
differences inflicted by uncertainties in OV during core He burning 
and by the occurrence of TP.


We conclude that the current uncertainties in the chemical profiles of
WDs, resulting from the progenitor evolution, impact the pulsational
period spectrum of ZZ Ceti stars. In this paper, we have quantified
for the first time the magnitude of this impact. The amount of the
resulting period changes is larger than the typical uncertainties of
the observed periods and also higher than the typical
accuracy of the period-to-period fits of 
  \citet{2012MNRAS.420.1462R} wich is, on average, less than 3.5
  s. However, we stress that the differences of periods we found
have to be considered as upper limits, since they correspond to
extreme uncertainties in prior evolution. In addition, the largest
period differences are found to occur only at certain specific modes,
and not at the whole period spectrum. Anyway, these uncertainties
should be taken into account in any asteroseismological analysis of ZZ
Ceti stars. We defer to a future work an assesment of this issue.

\begin{acknowledgements}
We acknowledge the valuable comments of our referee that
improved the original version of this paper. 
Part of this work was supported by AGENCIA through the Programa de
Modernizaci\'on Tecnol\'ogica BID 1728/OC-AR and by the PIP
112-200801-00940 grant from CONICET. This research has
made use of NASA Astrophysics Data System.
\end{acknowledgements}


\bibliographystyle{aa} 
\bibliography{paper-incertezas} 

\end{document}